\documentclass[twocolumn,aps,pra,superscriptaddress,longbibliography]{revtex4-1} 
\pdfoutput=1
\usepackage{graphicx} 
\usepackage[version=3]{mhchem} 
\usepackage[a4paper,plainpages=false,pdfstartview={XYZ null null 0.75}, colorlinks=true, citecolor=blue,linkcolor=blue]{hyperref}   
\usepackage[usenames,dvipsnames]{color}
\usepackage{amsfonts}
\usepackage{wrapfig}
\usepackage{subfigure}
\usepackage{amsmath}
\usepackage{gensymb}
\usepackage{url}
\usepackage{natbib}
\usepackage{cleveref}

\hypersetup{
    pdftitle = {Orbital angular momentum in electron diffraction and its use to determine chiral crystal symmetries},
    pdfkeywords = {Vortex beam, electron vortex,electron diffraction, orbital angular momentum, chiral, crystal}
}

\newcommand{\diff}{\ensuremath{\operatorname{d}}\!}

\newcommand{\nextline}[1][]{\nonumber\\#1&}

\newcommand{\Intbep}[3]{\ensuremath{\displaystyle{\int_{#1}^{#2}}\!\!\diff{#3}}\,}
\newcommand{\vt}[1]{\ensuremath{\boldsymbol{#1}}}

\graphicspath{{.}{Figures/}} 
\begin{document}
%\begin{frontmatter}

\title{Local orbital angular momentum revealed by spiral-phase-plate imaging in transmission-electron microscopy}

\author{Roeland Juchtmans}
\affiliation{EMAT, University of Antwerp, Groenenborgerlaan 171, 2020 Antwerp, Belgium}

\author{Jo Verbeeck}
\affiliation{EMAT, University of Antwerp, Groenenborgerlaan 171, 2020 Antwerp, Belgium}
\begin{abstract}
The orbital angular momentum (OAM) of light and matter waves is a parameter that has been getting increasingly more attention over the past couple of years. Beams with a well-defined OAM, the so-called vortex beams, are applied already in, e.g., telecommunication, astrophysics, nanomanipulation and chiral measurements in optics and electron microscopy. Also, the OAM of a wave induced by the interaction with a sample, has attracted a lot of interest. In all these experiments it is crucial to measure the exact (local) OAM content of the wave, whether it is an incoming vortex beam or an exit wave after interacting with a sample. In this work we investigate the use of spiral phase plates (SPPs) as an alternative to the programmable phase plates used in optics to measure OAM. We derive analytically how these can be used to study the local OAM components of any wave function. By means of numerical simulations we illustrate how the OAM of a pure vortex beam can be measured. We also look at a sum of misaligned vortex beams and show, by using SPPs, the position and the OAM of each individual beam  can be detected. Finally, we look at the OAM induced by a magnetic dipole on a free-electron wave and show how the SPP can be used to localize the magnetic poles and measure their ``magnetic charge.''\\
 Although our findings can be applied to study the OAM of any wave function, they are of particular interest for electron microscopy where versatile programmable phase plates do not yet exist.
\end{abstract}
\maketitle

%%%%%%%%%%%%%%%%%%%%%%
%Begin mainmatter
%%%%%%%%%%%%%%%%%%%%%%%
%
%

\section{Introduction}

Over the past 20 years orbital angular momentum (OAM) eigenstates in light and electron waves has been subject to intensive research. These eigenstates are characterized by a wavefunction $\Psi$ in cylindrical coordinates of the form
\begin{align}
\Psi(r,\phi,z)=\psi(r,z)e^{im\phi}, \label{VortexBeam}
\end{align}
where $\phi$ is the angular coordinate and the integer number $m$ is called the topological charge (TC).
Being eigenstates of the OAM operator they posses a well defined OAM of $m\hbar$ per particle.
Preceded by the early theoretical work of Nye and Berry \cite{Nye} and Allen et al. \cite{Allen}, numerous ways of creating OAM eigenstates have been designed in optics \cite{Vaughan1983,Bazhenov1990} and electron microscopy \cite{Uchida2010,Verbeeck2010,Clark2013,Beche2013,Beche2015}, acoustics \cite{Skeldon2008,Torabi2013}, and neutronics \cite{Clark2015}
In optics, these so-called vortex beams have found their way into a vast number of applications ranging from optical tweezers and nanomanipulation \cite{Galajda2001,Friese1998,Luo,He} and astrophysics \cite{Swartzlander2007,Serabyn2010,Berkhout2008} to telecommunication \cite{wangterabit,Roychowdhury2004}. 
%\cite{Molina-Terriza2007} 
Also, in electron microscopy, theoretical research and early experiments look very promising for vortex beams as a tool to manipulate nanocrystals \cite{Verbeeck2013} and investigate chiral crystals \cite{Juchtmans2015}. One of the most attractive among these applications is without a doubt  magnetic mapping on the atomic scale using the energy loss magnetic chiral dichroism signal from inelastic vortex scattering \cite{Schattschneider2014}.

For most of these applications, it is important to control the TC of the vortex beam and to create a beam that is a pure OAM eigenstate in stead of a mixture of several OAM eigenstates.
In order to look for the best setup to create high quality vortex beams, one has to be able to measure the TC with high accuracy.
With this purpose a variety of methods has been developed for optical beams \cite{Berkhout2009,Liu2015,Yongxin2011}, some from which an equivalent setup in electron microscopy was derived \cite{Guzzinati2014}. In optics, besides the total OAM of a beam, the local OAM density and OAM modes can also be measured by modal decomposition \cite{Dudley2012,Schulze2013} using computer generated holograms. 
This approach, however, is not applicable to electron vortex beams, since it requires versatile phase plates that do not yet exist in electron optics.

An interesting type of phase plate is the so-called spiral phase plates (SPP). 
Adding only a phase $e^{im\phi}$ and a TC $m$ to the wave in diffraction space, such phase plates can be created even for electron waves \cite{Beche2013,Beche2015}, where they have shown their use as an efficient way to create electron vortex beams.
%From previous work \cite{Juchtmans2015b} we know that when combined with an annular aperture selecting the first order Laue zone in reciprocal space, the chirality of a crystal can be seen on the atomic level.
In this paper we investigate the use of SPPs as an alternative to detect local OAM components in an electron wave. 
We analytically derive the action of an SPP and verify this with numerical simulations on Bessel beams. 
We explore the potential of this technique on examples of pure and superpositions of noncentered vortex beams. Finally we investigate how a magnetic dipole can be studied by using SPPs in an electron microscope.

\section{OAM-decomposition in real and reciprocal space}
Consider a two dimensional scalar wave function in polar coordinates, $\Psi(r,\phi)$. Taking the multipole series we write the function as a sum of OAM eigenstates of the form $e^{im\phi}$,
\begin{align}
\Psi(r,\phi)=\frac{1}{\sqrt{2\pi}}\sum_ma_m(r)e^{im\phi}.\label{OAMDecomposition}
\end{align}
The coefficients $a_m(r)$ are complex functions of the radial coordinate $r$ and are given by
\begin{align}
a_m(r)=\frac{1}{\sqrt{2\pi}}\Intbep{0}{2\pi}{\phi}\Psi(r,\phi)e^{-im\phi}\label{OAMCoefficients},
\end{align}
which can easily be verified by inserting \cref{OAMCoefficients} into \cref{OAMDecomposition}.

We can write the Fourier transform in polar coordinates as \cite{Wang}
\begin{align}
\tilde{\Psi}(k,\varphi)=& \frac{1}{\sqrt{2\pi}}\Intbep{0}{\infty}{r}r\Intbep{0}{2\pi}{\phi}\Psi(r,\phi)e^{ikr\cos(\phi-\varphi)}
\end{align}
By using the Jacobi--Anger identity
\begin{align}
e^{ikr\cos(\phi-\varphi)}=\sum_mi^{m}J_m(kr)e^{im(\phi-\varphi)},
\end{align}
we can rewrite $\tilde{\Psi}(k,\varphi)$ as a sum of OAM-eigenstates
\begin{align}
&\tilde{\Psi}(k,\varphi)\nonumber\\
&=\frac{1}{\sqrt{2\pi}}\Intbep{0}{\infty}{r}r\Intbep{0}{2\pi}{\phi}\Psi(r,\phi)\sum_mi^{m}J_m(k r)e^{im(\phi-\varphi)}\nextline =
\frac{1}{\sqrt{2\pi}}\Intbep{0}{\infty}{r}r\Intbep{0}{2\pi}{\phi}\Psi(r,\phi)\sum_mi^{-m}J_{-m}(k r)e^{im(\varphi-\phi)}\nextline =
\sum_m\tilde{a}_m(k)e^{im\varphi}.\label{FTOAMDecomposition}
\end{align}
The coefficients $\tilde{a}_m(k)$ and ${a}_m(k)$ are linked via the $m$th-order Hankel transform
\begin{align}
\tilde{a}_m(k)&=\frac{i^{m}}{\sqrt{2\pi}}\Intbep{0}{\infty}{r}r\Intbep{0}{2\pi}{\phi}\Psi(r,\phi)J_m(k r)e^{-im\phi}\nonumber\\
&=i^{m}\Intbep{0}{\infty}{r}rJ_m(kr)a_m(r)\label{atilde}\\
a_m(r)&=i^{-m}\Intbep{0}{\infty}{k}{k}J_m(kr)\tilde{a}_m(k).
\end{align}
In \cref{OAMDecomposition} and \cref{FTOAMDecomposition} we decomposed the function $\Psi(r,\phi)$ and its Fourier transform $\tilde{\Psi}(k,\varphi)$ in a basis of OAM-eigenstates. It is important to note that the coefficients $a_m(r)$ and $\tilde{a}_m(k)$ depend on the choice of origin around which the decomposition is made. Changing the origin will change the OAM components. We therefore introduce the notation $a^{\vt{R}}_m(\vt{r'})$, with $\vt{r'}=\vt{r}-\vt{R}$, to denote the OAM components for the decomposition around the point $\vt{R}$.
Let us illustrate this by investigating the OAM-decomposition of a Bessel beam with wave function
\begin{align}
\Psi(r,\phi)=J_{m'}(kr) e^{im'\phi},\label{eq:Besselbeam}
\end{align}
where $k$ is a parameter determining the width of the wave function. It is clear that the OAM coefficients of the beam are given by
\begin{align}
a_{m}(r)=J_{m'}(kr)\delta_{m',m}
\end{align}
when the decomposition is made with respect to the center of the beam. When shifting the beam over a distance $R$ in the $x$-direction however, the wave is given by the following expression
\begin{align}
\Psi'(r,\phi)=\Psi(\vt{r}+R\vt{e}_x)=\sum_{m'=-\infty}^{\infty}J_{1-m'}(kR)J_{m'}(kr)e^{im'\phi}.
\end{align}
Now the OAM components become
\begin{align}
a^R_m(r)=J_{1-m}(kR)J_m(kr),\label{eq:ShiftedBessel}
\end{align}
which shows that, indeed, the OAM components depend on the point $\vt{R}$ around which one makes the decomposition.\\

\section{Local detection of OAM components by using a spiral phase plate}
Measuring the OAM components of an optical or electron wave is not straightforward because these depend on both the amplitude and phase of the wave.
 Whereas the amplitude can be easily obtained from an image, the phase has to be retrieved by holographic reconstruction.
 For some applications one is more interested in the OAM-spectrum rather then the amplitude and phase of the entire wave and it would be more convenient to detect the OAM components directly.

\begin{figure}
	\includegraphics[width=\linewidth]{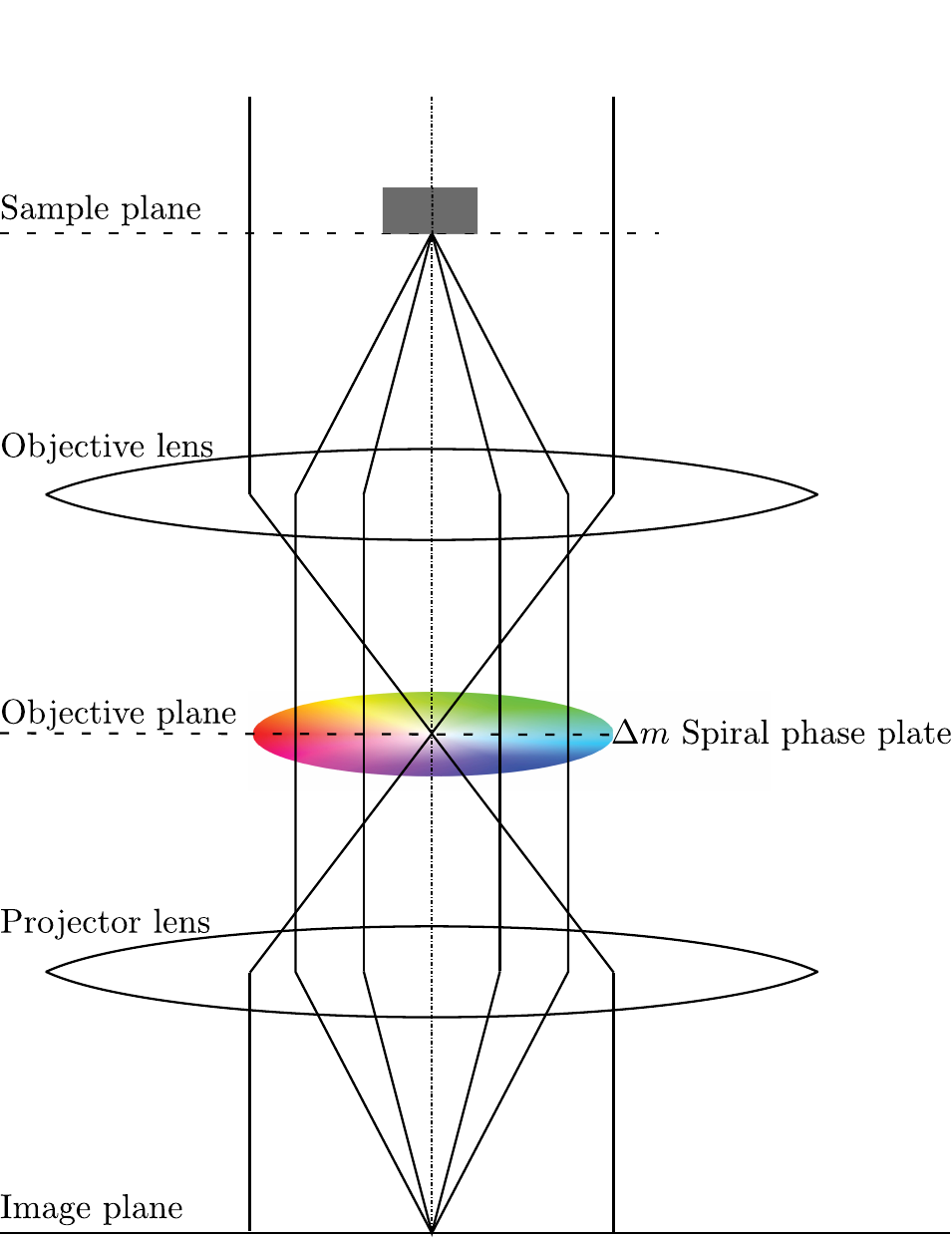}
	\caption{Sketch of the experimental setup to measure local OAM components of a wave function. In the back focal plane, a SPP is inserted adding a TC $\Delta m$ to the wave in Fourier space after which an image is taken. \label{fig:Setup}}
\end{figure}

With this goal in mind, let us study the effect of inserting a spiral phase plate (SPP) in the far-field plane of the wave of interest. These give the wave an extra angular-dependent phase factor $e^{i\Delta m\phi}$ to the wave in reciprocal space and adds a TC of $\Delta m$ to the wave, as illustrated in \cref{fig:Setup}.

\begin{figure}
	\includegraphics[width=\linewidth]{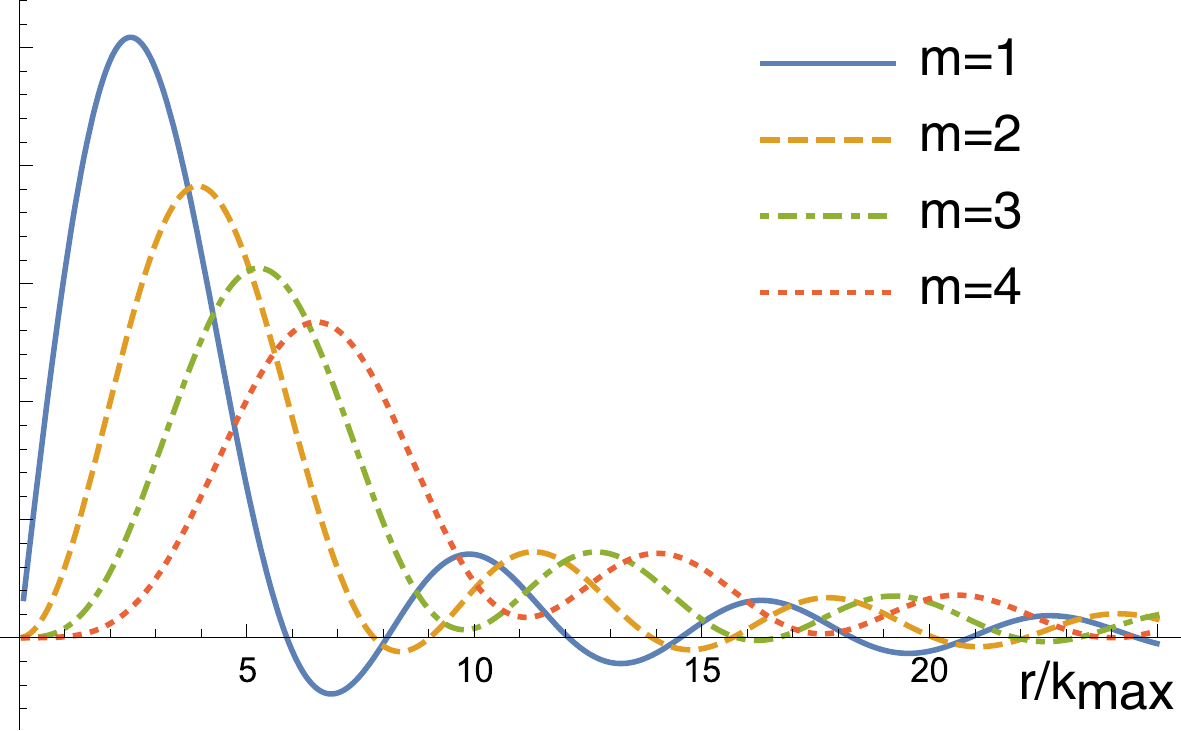}
	\caption{$T^{k_{max}}_{-\Delta m}(r)$ in arbitrary units, $r$ in units of $kmax^{-1}$.\label{fig:Tmrk}}
\end{figure}

The wave thus is multiplied in Fourier space  with a phase factor $e^{i\Delta m\varphi}$, and the multipole expression in \cref{FTOAMDecomposition} now becomes
\begin{align}
\tilde{\Psi}_{\Delta m}(k,\varphi)=&\sum_m\tilde{a}_m(k)e^{im\varphi}e^{i\Delta m\varphi}\nextline[=]
\sum_m\tilde{a}_{m-\Delta m}(k)e^{im\varphi}.
\end{align}
Going back to real space by taking the Fourier transform, we get
\begin{align}
&\Psi_{\Delta m}(r,\phi)=\mathcal{F}[\tilde{\Psi}_{\Delta m}](r,\phi)\nextline
=\Intbep{0}{\infty}{k}k\Intbep{0}{2\pi}{\varphi}\tilde{\Psi}_{\Delta m}(k_\perp,\varphi)e^{ik_\perp r\cos(\phi-\varphi)}\nextline 
=\Intbep{0}{\infty}{k}k\Intbep{0}{2\pi}{\varphi}\sum_{m,m'}\tilde{a}_{m-\Delta m}(k)e^{im\varphi}\nextline
\hphantom{=}\times i^{m'}J_{m'}(k r)e^{im'(\phi-\varphi)}\nextline
=2\pi i^m\Intbep{0}{\infty}{k}k\sum_{m}\tilde{a}_{m-\Delta m}(k) J_{m}(k r)e^{im\phi}
\end{align}
In general we can not simplify any further. However, looking at the point $r = 0$ we get using $J_0(0) = 1$ and $J_m(0) = 0$ for $m \neq 0$,
\begin{align}
\Psi_{\Delta m}(0)=2\pi \Intbep{0}{\infty}{k}k\tilde{a}_{-\Delta m}(k)
\end{align}
Filling in the expression for $\tilde{a}_m(k)$ in \cref{atilde}, gives
\begin{align}
\Psi_{\Delta m}(0)&=2\pi i^{-\Delta m}\Intbep{0}{\infty}{r}r a_{-\Delta m}(r)\Intbep{0}{\infty}{k}k J_{-\Delta m}(kr)\label{eq:PsiDeltam}
\end{align}
The integral over $k$ does not converge when integrating over the entire reciprocal space. However, in reality the upper boundary of the integral will be finite and determined by the maximum $k$ vector $k_{max}$, transmitted by the SPP, such that the expression becomes
\begin{align}
\Psi_{\Delta m}(0)&=2\pi i^{-\Delta m}\Intbep{0}{\infty}{r}r a_{-\Delta m}(r)\Intbep{0}{k_{max}}{k}k J_{-\Delta m}(kr)\nonumber\\
&=\Intbep{0}{\infty}{r}r a_{-\Delta m}(r)T^{k_{max}}_{-\Delta m}(r),\label{Psim0}
\end{align}
with
\begin{align}
T^{k_{max}}_{-\Delta m}(r)&=2\pi i^{-\Delta m}\Intbep{0}{k_{max}}{k}k J_{-\Delta m}(kr)\\
&=2\pi i^{-\Delta m}k^2_{max}\Intbep{0}{1}{k}k J_{-\Delta m}\left(\frac{kr}{k_{max}}\right).
\label{Kernel}
\end{align}
Eq. \ref{Kernel} is simply the radial part of the Fourier transform of a circular SPP with radius $k_{max}$ and TC $\Delta m$. Although at $r=0$ its value for $\Delta m\neq0$ goes to zero, with increasing $k_{max}$ the function becomes more and more condensed around $r=0$; see \cref{fig:Tmrk}. For $k_{max}\rightarrow\infty$ the radius of the Fourier transform of the SPP will go to zero and when it is small enough such that $a_{-\Delta m}(r)$ is approximately constant over this distance, we can approximate the wavefunction at the origin, \cref{Psim0}:
\begin{align}
\Psi_{\Delta m}(0)&=\Intbep{0}{\infty}{r}r a_{-\Delta m}(r)T^{k_{max}}_{-\Delta m}(r)\nonumber\\
&=\left(\lim_{r\rightarrow 0}a_{-\Delta m}(r)\right)\Intbep{0}{\infty}{r}r T^{k_{max}}_{-\Delta m}(r)\nonumber\\
&=K_{\Delta m} \lim_{r\rightarrow 0}a_{-\Delta m}(r),\label{Psim02}
\end{align}
with $K$ being a normalization constant. 
Here, care must be taken in taking the limit because $a_m(r)$ in \cref{OAMCoefficients} is not defined for $r=0$. 

The center of the recorded image with a $\Delta m$-SPP finally is given by the modulus squared of \cref{Psim02}

\begin{align}
I_{\Delta m}(0)=\left|\Psi_{\Delta m}(0)\right|^2= \left|K_{\Delta m} \lim_{r\rightarrow 0}a_{-\Delta m}(r)\right|^2,\label{Im0}
\end{align}

\cref{Im0} shows that the central point in the image (i.e., the point on the optical axis) only depends on $ \lim_{r\rightarrow 0}a_{-\Delta m}(r)$, when the decomposition is made around the point on the optical axis. 

\begin{figure}[t]
    \includegraphics[width=.7\linewidth]{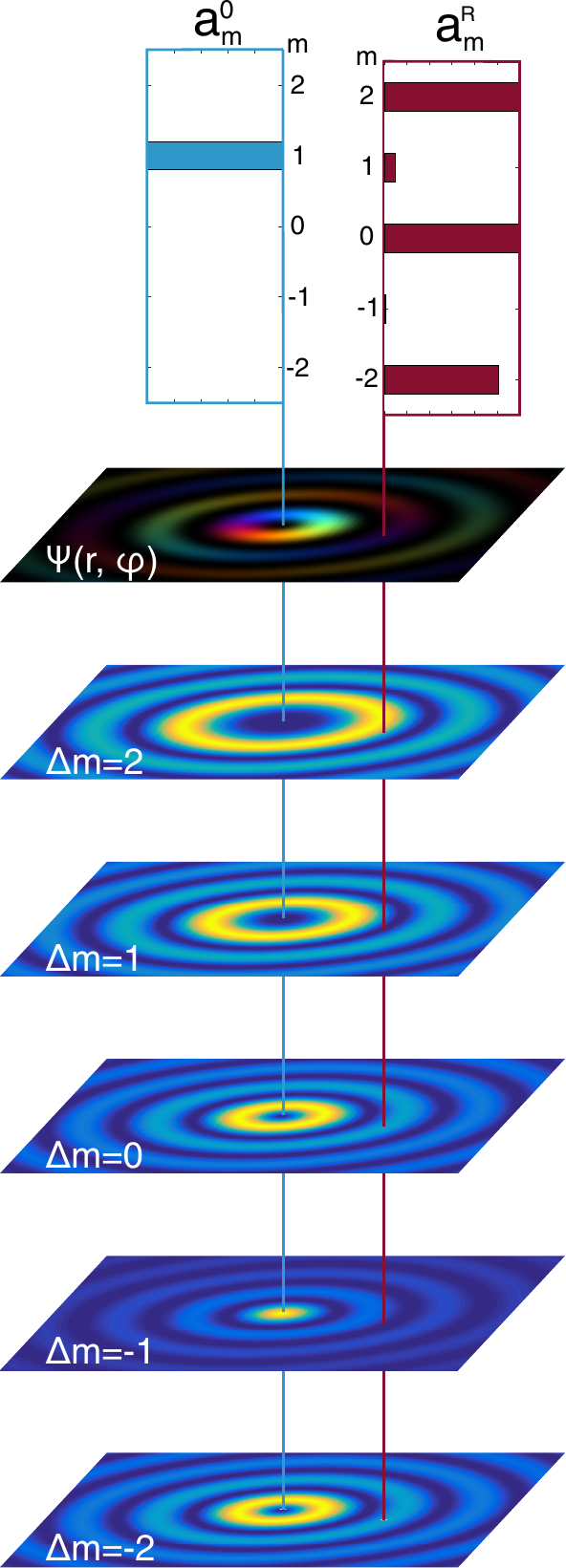}
    \caption{Illustration of how the intensity in each point of a SPP image is linked to the local OAM component of the wave function. The upper image shows the amplitude and phase of a Bessel beam with TC $m=1$; see \cref{eq:Besselbeam}. In the center of the beam, the only nonzero OAM component is $a_1$ and the intensity in the center will be zero in each but the $\Delta m=-1$ SPP image. In any other point of the wave, all components can be present and the intensity in the SPP images varies as such. \label{fig:BesselFlags}}
\end{figure}

Up to this point, we only looked at the center of the image. 
 When we shift the wave of interest in real space, we put a different point on the optical axis of the microscope and the intensity in the center of the new image then is only determined by the  $-\Delta m$th component of the decomposition around this new point.
However, the only effect of a shift of the wave in real space is a shift of the image.
Therefore \textbf{every} point in the image is determined only by the  $-\Delta m$th component of the expansion at the corresponding point of the wave in real space, $\lim_{r\rightarrow 0}a^{\vt{R}}_{-\Delta m}(r)$
\begin{align}
I_{\Delta m}(\vt{R})&= \left|K_{\Delta m} \lim_{r\rightarrow 0}a^{\vt{R}}_{-\Delta m}(r)\right|^2\nonumber\\
&= \left|K_{\Delta m}a^{\vt{R}}_{-\Delta m}\right|,\label{Imr}
\end{align}
where we call 
\begin{align}
a^{\vt{R}}_{-\Delta m}=\lim_{r\rightarrow 0}a^{\vt{R}}_{-\Delta m}(r)\label{eq:localOAM}
\end{align}
the local OAM components of the wave.
This is illustrated in \cref{fig:BesselFlags}, where we compare $a^{\vt{R}}_{-\Delta m}$ in two points of the wave function of a Bessel beam, \cref{eq:Besselbeam}, with the intensity in the corresponding point in the SPP images with different $\Delta m$.

These results explain nicely the observations made by F\"urhapter et al. \cite{Furhapter2005}, where they use SPP images to enhance the contrast at the edges of phase objects.
For regions in the exitwave where the amplitude and phase are constant, the only nonzero OAM component is the $a^{\vt{R}}_{0}$ component, and inserting a SPP will only show regions where the phase or amplitude of the wave is changing, such as at the edge of phase objects.

\section{Application: Measuring OAM of vortex beams}
An obvious application of the SPP is the determination of the TC of pure vortex beams. Since their wave function is defined as in \cref{VortexBeam}, the OAM-decomposition around the center of the beam only has one component, $a_m(r)$, with $m$ being the TC of the beam. The TC can be determined from images using SPPs with different TC $\Delta m$, where the center of the beam will be dark in every image but the one with SPP $\Delta m =-m$. In \cref{fig:BesselFlags} this is simulated for a Bessel beam with TC $m=1$. 
In Fourier space the phase in the central pixel is not defined and we therefore put it equal to zero.
In optics a wide range of methods are developed to measure the OAM of pure vortex beams by using programmable phase plates. In electron microscopy on the other hand, such programmable phase plates are missing which makes those methods impractical. A SPP with arbitrary $\Delta m$, however, can be made by using magnetized needles \cite{Beche2013}, making the method described here of great future potential.

Knowing every point to be determined by the local OAM components, let us now look at a slightly more complicated situation of a superposition of two Bessel beams with opposite TC and misaligned vortex cores, e.g., the sum of two shifted Bessel beams with TC $m=\pm 1$ of which the intensity looks like Fig. 4(b). A SPP with $\Delta m= -1$ enables us to localize the center of the vortex beam with TC $m=1$; see Fig. 4(c). Because the $m=1$-component will be highest around the center of this vortex beam, it will light up in the $\Delta m=-1$ SPP image. The opposite SPP localizes the other vortex core.

\begin{figure}[htb]
	\includegraphics[width=.9\linewidth]{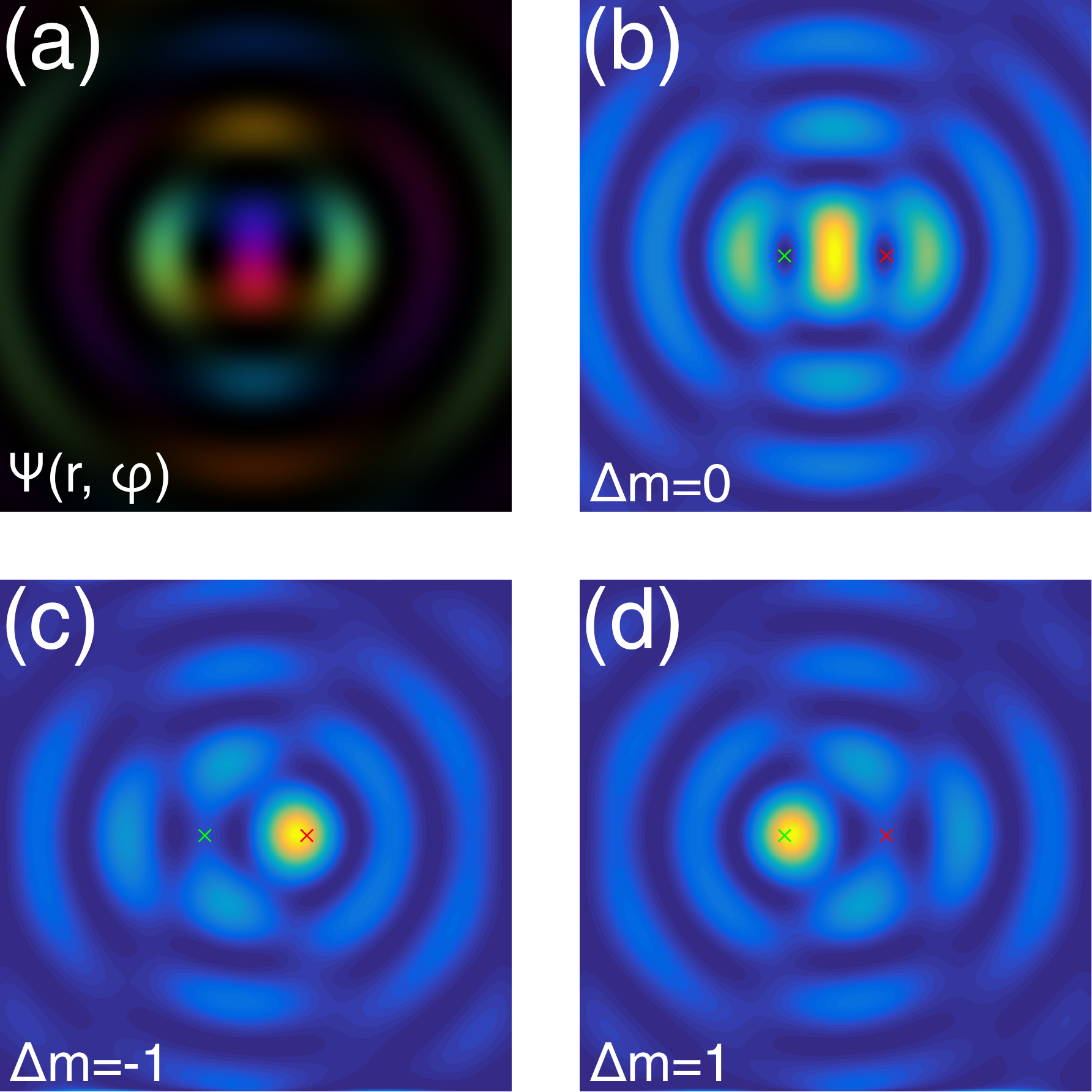}
	\caption{Sum of two Bessel beams with TC $m=-1$ positioned at the green X and $m=1$ positioned at the red X. (a) Phase plot-amplitude plot. (b) Intensity plot of panel (a). (c) Image obtained with SPP with TC $\Delta m=-1$ showing the position of the $m=1$ Bessel beam. (d) Image obtained with SPP with TC $\Delta m=1$ indicating the position of the $m=-1$ Bessel beam.  \label{fig:Bessels}}
\end{figure}

Despite the maximum intensity in the images being slightly shifted with respect to the location of the vortex cores because of interference with the other Bessel beam, the modal OAM decomposition allows us to clearly identify the two opposite vortex cores in a convenient way.

\section{Application: Imaging a magnetic dipole}
\begin{figure}[thb]
	\includegraphics[width=.9\linewidth]{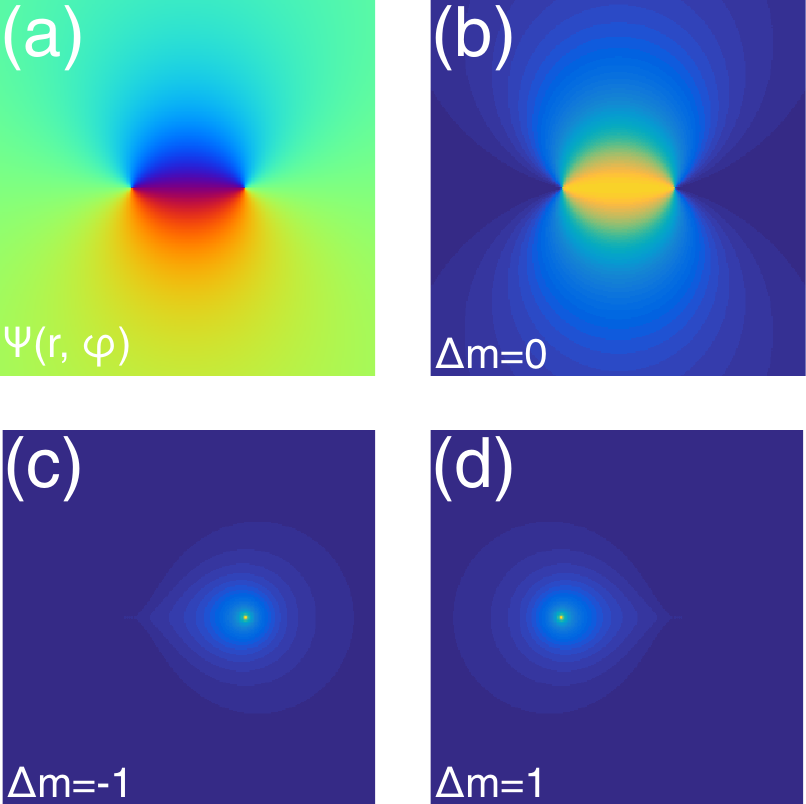}
	\caption{(a) Aharonov--Bohm-induced phase map of a plane wave passing a magnetic dipole. (b) Image obtained without SPP, only blocking central pixel. (c) and (d) Images obtained with $\Delta m=-1$ and $\Delta m=1$ phase plate, respectively, from which the ``magnetic charge'' of the poles can be deduced. \label{fig:MagneticDipole}}
\end{figure}

First noted in ref. \cite{Tonomura1987}, the Aharonov--Bohm phase caused by a magnetic monopole transforms a passing plane electron wave into a vortex electron wave. In fact, it is this effect that B\'ech\'e et al. \cite{Beche2013} used to generate electron vortex beams near the tip of a single-domain magnetized needle.\\
Following this principle, a magnetic dipole creates two phase singularities in the plane wave electron at the position of its two poles with two opposite TC's depending on the strength of the magnetic flux inside the needle; see \cref{fig:MagneticDipole}.a. 
Here, two vortices with TC $m\pm1$ are created at the magnetic poles.

Because the magnetic field only causes a phase shift in the plane-wave electron, we can not determine the ``magnetic charge'' of the poles by looking at the intensity of the electron wave in real space. However, if we apply the OAM decomposition proposed in this paper, \cref{fig:MagneticDipole}.c and \cref{fig:MagneticDipole}.d clearly reveal the position of both ends separately.

Note that, although the magnetic dipole only causes a phase shift, we do observe intensity in the $\Delta m=0$ image. This effect is due to the way we choose to show the simulations. For the $\Delta m\neq0$ SPP images, the phase of the phase plate in the center of reciprocal space is ill-defined. Therefore we put the central pixel equal to zero. For comparison with the $m\neq0$ SPP images, in the $\Delta m =0$ image we also put the central pixel in Fourier space equal to zero. This is equivalent with a dark field image which causes the phase shift to become visible. However, the polarity of both poles remains undetectable without a SPP and the contrast is low.
\section{Discussion}
Inserting a $\Delta m$ SPP in the back focal plane of a microscope allows us to look at the local OAM component $a_{-\Delta m}^{\vt{R}}$ defined in \cref{eq:localOAM}. As can be seen in \cref{fig:BesselFlags}, this can be used to determine the OAM of a pure vortex beam, where the intensity at the center of the beam is zero except for the $-m$ SPP image. The local character of the OAM-measurement with SPPs allows us to localize the center of individual vortex beams when looking at a sum of misaligned vortex beams with different OAM; see \cref{fig:Bessels}. We can extend this to look at the signature of certain magnetic fields imposed on a free-electron wave. Looking with a SPP at an electron wave after interaction with a magnetic dipole, for instance, clearly localizes the position of the individual poles as well as their ``magnetic charge''.
All setups looked at in this work, require multiple SPPs that add different amounts of TC to the electron wave in diffraction space. This is done preferentiallyby one single SPP where the TC can be chosen freely. Although this has not been achieved yet, which is why no experiments are included in this paper, significant progress has been made recently in this field. B\'ech\'e et al \cite{Beche2013} showed that the tip of a magnetized needle centered on a circular aperture, generates a SPP with OAM determined by the magnetic flux inside the needle. In ref. \cite{Verbeeck2015} the same authors demonstrated a SPP where the magnetization of the needle is changed by the external field of a macroscopic solenoid around it, thus changing the sign of the TC in microsecond intervals. By replacing the magnetic needle with a microscopic solenoid, a fully dynamical SPP could be created where the TC can take any value by controlling the current running through the coil. Note that we do not consider this a fully programmable phase plate, because this would require full control of the phase in any point of the wave separately.
Another possibility is to insert a so-called fork aperture \cite{Verbeeck2010} into the back focal plane of the microscope, which should also give a series of $\Delta m$ images spaced apart.

\section{Conclusion}
We investigated the use of spiral phase plates (SPPs) as a tool to measure the local orbital angular momentum (OAM) components of photons, electrons, or any other wave function. We found that inserting a SPP that adds a phase factor $e^{i\Delta m \varphi}$ to the wave in the back focal plane gives a real-space image in $I(\vt{R})$ that is proportional to $a^{\vt{R}}_{-\Delta m}$, the $\Delta m$th local OAM component around the point $\vt{R}$; see \cref{Imr}. 

Not only does this explain the observations made by F\"{u}rhapter et al.\cite{Furhapter2005}, where a SPP visualizes the edges of phase objects in electron microscopy, it also enables us to study OAM components on the local level in a similar way to the OAM-mode decomposition using computer generated holograms as described by  Dudley et al. \cite{Dudley2012} and Schulze et al. \cite{Schulze2013}. The main advantage of the SPP method, however, is that it can also be applied to systems for which programmable phase plates are not available as in, e.g., electron microscopes.

We demonstrated by means of numerical simulations how this setup can be used to determine the OAM of pure electron vortex beams where the intensity in the center of a beam with topological charge (TC) $m$ is only nonzero by using a $-m$ SPP. Also for beams that are a sum of misaligned vortex beams with different TCs, the center of each vortex and its corresponding TC can be determined from a series of images with different SPPs. In a last example we showed a hint of the potential use of SPPs in electron microscopes to study magnetic materials by simulating the SPP images for a plane wave passing a magnetic dipole that adds local OAM to the electron. By using our method, the position of the magnetic poles as well as their ``magnetic charg'' can be determined. 
Although fully controllable phase plates do not yet exist in electron microscopy, the method described in this paper only requires the use of SPPs. These phase plates already are made bu using magnetic ``monopole-like'' fields at the tip of a magnetized needle \cite{Beche2013} of which the sign of the topological charge can be changed in only a few microseconds \cite{Verbeeck2015}. Full control of the TC could be achieved by replacing the magnetic needle by a microscopic solenoid or a so-called fork aperture \cite{Verbeeck2010}.

\section{acknowledgments}
The authors acknowledge support from the FWO (Aspirant Fonds Wetenschappelijk Onderzoek--Vlaanderen), the EU under the Seventh Framework Program (FP7) under a contract for an Integrated Infrastructure Initiative, Reference No. 312483-ESTEEM2 and ERC Starting Grant 278510 VORTEX.


\begin{thebibliography}{34}%
	\makeatletter
	\providecommand \@ifxundefined [1]{%
		\@ifx{#1\undefined}
	}%
	\providecommand \@ifnum [1]{%
		\ifnum #1\expandafter \@firstoftwo
		\else \expandafter \@secondoftwo
		\fi
	}%
	\providecommand \@ifx [1]{%
		\ifx #1\expandafter \@firstoftwo
		\else \expandafter \@secondoftwo
		\fi
	}%
	\providecommand \natexlab [1]{#1}%
	\providecommand \enquote  [1]{``#1''}%
	\providecommand \bibnamefont  [1]{#1}%
	\providecommand \bibfnamefont [1]{#1}%
	\providecommand \citenamefont [1]{#1}%
	\providecommand \href@noop [0]{\@secondoftwo}%
	\providecommand \href [0]{\begingroup \@sanitize@url \@href}%
	\providecommand \@href[1]{\@@startlink{#1}\@@href}%
	\providecommand \@@href[1]{\endgroup#1\@@endlink}%
	\providecommand \@sanitize@url [0]{\catcode `\\12\catcode `\$12\catcode
		`\&12\catcode `\#12\catcode `\^12\catcode `\_12\catcode `\%12\relax}%
	\providecommand \@@startlink[1]{}%
	\providecommand \@@endlink[0]{}%
	\providecommand \url  [0]{\begingroup\@sanitize@url \@url }%
	\providecommand \@url [1]{\endgroup\@href {#1}{\urlprefix }}%
	\providecommand \urlprefix  [0]{URL }%
	\providecommand \Eprint [0]{\href }%
	\providecommand \doibase [0]{http://dx.doi.org/}%
	\providecommand \selectlanguage [0]{\@gobble}%
	\providecommand \bibinfo  [0]{\@secondoftwo}%
	\providecommand \bibfield  [0]{\@secondoftwo}%
	\providecommand \translation [1]{[#1]}%
	\providecommand \BibitemOpen [0]{}%
	\providecommand \bibitemStop [0]{}%
	\providecommand \bibitemNoStop [0]{.\EOS\space}%
	\providecommand \EOS [0]{\spacefactor3000\relax}%
	\providecommand \BibitemShut  [1]{\csname bibitem#1\endcsname}%
	\let\auto@bib@innerbib\@empty
	%</preamble>
	\bibitem [{\citenamefont {Nye}\ and\ \citenamefont {Berry}(1974)}]{Nye}%
	\BibitemOpen
	\bibfield  {author} {\bibinfo {author} {\bibfnamefont {J.~F.}\ \bibnamefont
			{Nye}}\ and\ \bibinfo {author} {\bibfnamefont {M.~V.}\ \bibnamefont
			{Berry}},\ }\bibfield  {title} {\enquote {\bibinfo {title} {Dislocations in
				wave trains},}\ }\href {\doibase 10.1098/rspa.1974.0012} {\bibfield
		{journal} {\bibinfo  {journal} {Proc. R. Soc. London A}\ }\textbf {\bibinfo
			{volume} {336}},\ \bibinfo {pages} {165--190} (\bibinfo {year}
		{1974})}\BibitemShut {NoStop}%
	\bibitem [{\citenamefont {Allen}\ \emph {et~al.}(1992)\citenamefont {Allen},
		\citenamefont {Beijersbergen}, \citenamefont {Spreeuw},\ and\ \citenamefont
		{Woerdman}}]{Allen}%
	\BibitemOpen
	\bibfield  {author} {\bibinfo {author} {\bibfnamefont {L.}~\bibnamefont
			{Allen}}, \bibinfo {author} {\bibfnamefont {M.~W.}\ \bibnamefont
			{Beijersbergen}}, \bibinfo {author} {\bibfnamefont {R.~J.~C.}\ \bibnamefont
			{Spreeuw}}, \ and\ \bibinfo {author} {\bibfnamefont {J.~P.}\ \bibnamefont
			{Woerdman}},\ }\bibfield  {title} {\enquote {\bibinfo {title} {Orbital
				angular momentum of light and the transformation of laguerre-gaussian laser
				modes},}\ }\href {\doibase 10.1103/PhysRevA.45.8185} {\bibfield  {journal}
		{\bibinfo  {journal} {Phys. Rev. A}\ }\textbf {\bibinfo {volume} {45}},\
		\bibinfo {pages} {8185--8189} (\bibinfo {year} {1992})}\BibitemShut {NoStop}%
	\bibitem [{\citenamefont {Vaughan}\ and\ \citenamefont
		{Willetts}(1983)}]{Vaughan1983}%
	\BibitemOpen
	\bibfield  {author} {\bibinfo {author} {\bibfnamefont {J.~M.}\ \bibnamefont
			{Vaughan}}\ and\ \bibinfo {author} {\bibfnamefont {D.~V.}\ \bibnamefont
			{Willetts}},\ }\bibfield  {title} {\enquote {\bibinfo {title} {Temporal and
				interference fringe analysis of tem01$\ast$ laser modes},}\ }\href {\doibase
		10.1364/JOSA.73.001018} {\bibfield  {journal} {\bibinfo  {journal} {J. Opt.
				Soc. Am.}\ }\textbf {\bibinfo {volume} {73}},\ \bibinfo {pages} {1018--1021}
		(\bibinfo {year} {1983})}\BibitemShut {NoStop}%
	\bibitem [{\citenamefont {Bazhenov}\ \emph {et~al.}(1990)\citenamefont
		{Bazhenov}, \citenamefont {Vasnetsov},\ and\ \citenamefont
		{Soskin}}]{Bazhenov1990}%
	\BibitemOpen
	\bibfield  {author} {\bibinfo {author} {\bibfnamefont {VY}~\bibnamefont
			{Bazhenov}}, \bibinfo {author} {\bibfnamefont {MV}~\bibnamefont {Vasnetsov}},
		\ and\ \bibinfo {author} {\bibfnamefont {MS}~\bibnamefont {Soskin}},\
	}\bibfield  {title} {\enquote {\bibinfo {title} {{Laser beams with screw
				dislocations in their wavefronts}},}\ }\href
{http://195.178.214.34/ps/1159/article\_17529.pdf} {\bibfield  {journal}
	{\bibinfo  {journal} {Jetp. Lett.}\ }\textbf {\bibinfo {volume} {52}},\
	\bibinfo {pages} {429--431} (\bibinfo {year} {1990})}\BibitemShut {NoStop}%
\bibitem [{\citenamefont {Uchida}\ and\ \citenamefont
	{Tonomura}(2010)}]{Uchida2010}%
\BibitemOpen
\bibfield  {author} {\bibinfo {author} {\bibfnamefont {Masaya}\ \bibnamefont
		{Uchida}}\ and\ \bibinfo {author} {\bibfnamefont {Akira}\ \bibnamefont
		{Tonomura}},\ }\bibfield  {title} {\enquote {\bibinfo {title} {{Generation of
				electron beams carrying orbital angular momentum.}}}\ }\href {\doibase
	10.1038/nature08904} {\bibfield  {journal} {\bibinfo  {journal} {Nature}\
	}\textbf {\bibinfo {volume} {464}},\ \bibinfo {pages} {737--739} (\bibinfo
	{year} {2010})}\BibitemShut {NoStop}%
\bibitem [{\citenamefont {Verbeeck}\ \emph {et~al.}(2010)\citenamefont
	{Verbeeck}, \citenamefont {Tian},\ and\ \citenamefont
	{Schattschneider}}]{Verbeeck2010}%
\BibitemOpen
\bibfield  {author} {\bibinfo {author} {\bibfnamefont {J}~\bibnamefont
		{Verbeeck}}, \bibinfo {author} {\bibfnamefont {H}~\bibnamefont {Tian}}, \
	and\ \bibinfo {author} {\bibfnamefont {P}~\bibnamefont {Schattschneider}},\
}\bibfield  {title} {\enquote {\bibinfo {title} {{Production and application
			of electron vortex beams.}}}\ }\href {\doibase 10.1038/nature09366}
{\bibfield  {journal} {\bibinfo  {journal} {Nature}\ }\textbf {\bibinfo
		{volume} {467}},\ \bibinfo {pages} {301--4} (\bibinfo {year}
	{2010})}\BibitemShut {NoStop}%
\bibitem [{\citenamefont {Clark}\ \emph {et~al.}(2013)\citenamefont {Clark},
	\citenamefont {B\'{e}ch\'{e}}, \citenamefont {Guzzinati}, \citenamefont
	{Lubk}, \citenamefont {Mazilu}, \citenamefont {{Van Boxem}},\ and\
	\citenamefont {Verbeeck}}]{Clark2013}%
\BibitemOpen
\bibfield  {author} {\bibinfo {author} {\bibfnamefont {L.}~\bibnamefont
		{Clark}}, \bibinfo {author} {\bibfnamefont {A.}~\bibnamefont
		{B\'{e}ch\'{e}}}, \bibinfo {author} {\bibfnamefont {G.}~\bibnamefont
		{Guzzinati}}, \bibinfo {author} {\bibfnamefont {A.}~\bibnamefont {Lubk}},
	\bibinfo {author} {\bibfnamefont {M.}~\bibnamefont {Mazilu}}, \bibinfo
	{author} {\bibfnamefont {R.}~\bibnamefont {{Van Boxem}}}, \ and\ \bibinfo
	{author} {\bibfnamefont {J.}~\bibnamefont {Verbeeck}},\ }\bibfield  {title}
{\enquote {\bibinfo {title} {{Exploiting Lens Aberrations to Create
				Electron-Vortex Beams}},}\ }\href
{http://link.aps.org/doi/10.1103/PhysRevLett.111.064801} {\bibfield
	{journal} {\bibinfo  {journal} {Phys. Rev. Lett.}\ }\textbf {\bibinfo
		{volume} {111}},\ \bibinfo {pages} {064801} (\bibinfo {year}
	{2013})}\BibitemShut {NoStop}%
\bibitem [{\citenamefont {B\'{e}ch\'{e}}\ \emph {et~al.}(2014)\citenamefont
	{B\'{e}ch\'{e}}, \citenamefont {{Van Boxem}}, \citenamefont {{Van
			Tendeloo}},\ and\ \citenamefont {Verbeeck}}]{Beche2013}%
\BibitemOpen
\bibfield  {author} {\bibinfo {author} {\bibfnamefont {Armand}\ \bibnamefont
		{B\'{e}ch\'{e}}}, \bibinfo {author} {\bibfnamefont {Ruben}\ \bibnamefont
		{{Van Boxem}}}, \bibinfo {author} {\bibfnamefont {Gustaaf}\ \bibnamefont
		{{Van Tendeloo}}}, \ and\ \bibinfo {author} {\bibfnamefont {Jo}~\bibnamefont
		{Verbeeck}},\ }\bibfield  {title} {\enquote {\bibinfo {title} {{Magnetic
				monopole field exposed by electrons}},}\ }\href {\doibase 10.1038/nphys2816}
{\bibfield  {journal} {\bibinfo  {journal} {Nature Phys.}\ }\textbf {\bibinfo
		{volume} {10}},\ \bibinfo {pages} {26--29} (\bibinfo {year}
	{2014})}\BibitemShut {NoStop}%
\bibitem [{\citenamefont {B\'{e}ch\'{e}}\ \emph {et~al.}()\citenamefont
	{B\'{e}ch\'{e}}, \citenamefont {Winkler}, \citenamefont {H.}, \citenamefont
	{F.},\ and\ \citenamefont {J.}}]{Beche2015}%
\BibitemOpen
\bibfield  {author} {\bibinfo {author} {\bibfnamefont {Armand}\ \bibnamefont
		{B\'{e}ch\'{e}}}, \bibinfo {author} {\bibfnamefont {R.}~\bibnamefont
		{Winkler}}, \bibinfo {author} {\bibfnamefont {Plank}\ \bibnamefont {H.}},
	\bibinfo {author} {\bibfnamefont {Hofer}\ \bibnamefont {F.}}, \ and\ \bibinfo
	{author} {\bibfnamefont {Verbeeck}\ \bibnamefont {J.}},\ }\bibfield  {title}
{\enquote {\bibinfo {title} {{Focus electron beam induced deposition as a
				tool to create electron vortices}},}\ }\href@noop {} {\bibinfo  {journal}
	{Micron}\ }\BibitemShut {NoStop}%
\bibitem [{\citenamefont {Skeldon}\ \emph {et~al.}(2008)\citenamefont
	{Skeldon}, \citenamefont {Wilson}, \citenamefont {Edgar},\ and\ \citenamefont
	{Padgett}}]{Skeldon2008}%
\BibitemOpen
\bibfield  {journal} {  }\bibfield  {author} {\bibinfo {author} {\bibfnamefont
		{K~D}\ \bibnamefont {Skeldon}}, \bibinfo {author} {\bibfnamefont
		{C}~\bibnamefont {Wilson}}, \bibinfo {author} {\bibfnamefont {M}~\bibnamefont
		{Edgar}}, \ and\ \bibinfo {author} {\bibfnamefont {M~J}\ \bibnamefont
		{Padgett}},\ }\bibfield  {title} {\enquote {\bibinfo {title} {{An acoustic
				spanner and its associated rotational Doppler shift}},}\ }\href {\doibase
	10.1088/1367-2630/10/1/013018} {\bibfield  {journal} {\bibinfo  {journal}
		{New Journal of Physics}\ }\textbf {\bibinfo {volume} {10}},\ \bibinfo
	{pages} {013018} (\bibinfo {year} {2008})}\BibitemShut {NoStop}%
\bibitem [{\citenamefont {Torabi}\ and\ \citenamefont
	{Rezaei}(2013)}]{Torabi2013}%
\BibitemOpen
\bibfield  {author} {\bibinfo {author} {\bibfnamefont {Reza}\ \bibnamefont
		{Torabi}}\ and\ \bibinfo {author} {\bibfnamefont {Zahra}\ \bibnamefont
		{Rezaei}},\ }\bibfield  {title} {\enquote {\bibinfo {title} {{The effect of
				Dirac phase on acoustic vortex in media with screw dislocation}},}\ }\href
{\doibase 10.1016/j.physleta.2013.05.014} {\bibfield  {journal} {\bibinfo
		{journal} {Physics Letters, Section A: General, Atomic and Solid State
			Physics}\ }\textbf {\bibinfo {volume} {377}},\ \bibinfo {pages} {1668--1671}
	(\bibinfo {year} {2013})},\ \Eprint {http://arxiv.org/abs/arXiv:1306.3038v1}
{arXiv:arXiv:1306.3038v1} \BibitemShut {NoStop}%
\bibitem [{\citenamefont {Clark}\ \emph {et~al.}(2015)\citenamefont {Clark},
	\citenamefont {Barankov}, \citenamefont {Huber}, \citenamefont {Arif},
	\citenamefont {Cory},\ and\ \citenamefont {Pushin}}]{Clark2015}%
\BibitemOpen
\bibfield  {author} {\bibinfo {author} {\bibfnamefont {Charles~W}\
		\bibnamefont {Clark}}, \bibinfo {author} {\bibfnamefont {Roman}\ \bibnamefont
		{Barankov}}, \bibinfo {author} {\bibfnamefont {Michael~G}\ \bibnamefont
		{Huber}}, \bibinfo {author} {\bibfnamefont {Muhammad}\ \bibnamefont {Arif}},
	\bibinfo {author} {\bibfnamefont {David~G}\ \bibnamefont {Cory}}, \ and\
	\bibinfo {author} {\bibfnamefont {Dmitry~A}\ \bibnamefont {Pushin}},\
}\bibfield  {title} {\enquote {\bibinfo {title} {{Controlling neutron orbital
			angular momentum.}}}\ }\href {\doibase 10.1038/nature15265} {\bibfield
{journal} {\bibinfo  {journal} {Nature}\ }\textbf {\bibinfo {volume} {525}},\
\bibinfo {pages} {504--6} (\bibinfo {year} {2015})}\BibitemShut {NoStop}%
\bibitem [{\citenamefont {Galajda}\ and\ \citenamefont
	{Ormos}(2001)}]{Galajda2001}%
\BibitemOpen
\bibfield  {author} {\bibinfo {author} {\bibfnamefont {P\'{e}ter}\
		\bibnamefont {Galajda}}\ and\ \bibinfo {author} {\bibfnamefont {P\`{a}l}\
		\bibnamefont {Ormos}},\ }\bibfield  {title} {\enquote {\bibinfo {title}
		{Complex micromachines produced and driven by light},}\ }\href {\doibase
	http://dx.doi.org/10.1063/1.1339258} {\bibfield  {journal} {\bibinfo
		{journal} {Applied Physics Letters}\ }\textbf {\bibinfo {volume} {78}},\
	\bibinfo {pages} {249--251} (\bibinfo {year} {2001})}\BibitemShut {NoStop}%
\bibitem [{\citenamefont {Friese}\ \emph {et~al.}(1998)\citenamefont {Friese},
	\citenamefont {Nieminen}, \citenamefont {Heckenberg},\ and\ \citenamefont
	{Rubinsztein-Dunlop}}]{Friese1998}%
\BibitemOpen
\bibfield  {author} {\bibinfo {author} {\bibfnamefont {M.~E.~J.}\
		\bibnamefont {Friese}}, \bibinfo {author} {\bibfnamefont {T.~A.}\
		\bibnamefont {Nieminen}}, \bibinfo {author} {\bibfnamefont {N.~R.}\
		\bibnamefont {Heckenberg}}, \ and\ \bibinfo {author} {\bibfnamefont
		{H.}~\bibnamefont {Rubinsztein-Dunlop}},\ }\bibfield  {title} {\enquote
	{\bibinfo {title} {{Optical alignment and spinning of laser-trapped
				microscopic particles}},}\ }\href {\doibase 10.1038/28566} {\bibfield
	{journal} {\bibinfo  {journal} {Nature}\ }\textbf {\bibinfo {volume} {394}},\
	\bibinfo {pages} {348--350} (\bibinfo {year} {1998})}\BibitemShut {NoStop}%
\bibitem [{\citenamefont {Luo}\ \emph {et~al.}(2000)\citenamefont {Luo},
	\citenamefont {Sun},\ and\ \citenamefont {An}}]{Luo}%
\BibitemOpen
\bibfield  {author} {\bibinfo {author} {\bibfnamefont {Zong-Ping}\
		\bibnamefont {Luo}}, \bibinfo {author} {\bibfnamefont {Yu-Long}\ \bibnamefont
		{Sun}}, \ and\ \bibinfo {author} {\bibfnamefont {Kai-Nan}\ \bibnamefont
		{An}},\ }\bibfield  {title} {\enquote {\bibinfo {title} {An optical spin
			micromotor},}\ }\href
{http://scitation.aip.org/content/aip/journal/apl/76/13/10.1063/1.126165}
{\bibfield  {journal} {\bibinfo  {journal} {Appl. Phys. Lett.}\ }\textbf
	{\bibinfo {volume} {76}},\ \bibinfo {pages} {1779--1781} (\bibinfo {year}
	{2000})}\BibitemShut {NoStop}%
\bibitem [{\citenamefont {He}\ \emph {et~al.}(1995)\citenamefont {He},
	\citenamefont {Friese}, \citenamefont {Heckenberg},\ and\ \citenamefont
	{Rubinsztein-Dunlop}}]{He}%
\BibitemOpen
\bibfield  {author} {\bibinfo {author} {\bibfnamefont {H.}~\bibnamefont
		{He}}, \bibinfo {author} {\bibfnamefont {M.~E.~J.}\ \bibnamefont {Friese}},
	\bibinfo {author} {\bibfnamefont {N.~R.}\ \bibnamefont {Heckenberg}}, \ and\
	\bibinfo {author} {\bibfnamefont {H.}~\bibnamefont {Rubinsztein-Dunlop}},\
}\bibfield  {title} {\enquote {\bibinfo {title} {Direct observation of
		transfer of angular momentum to absorptive particles from a laser beam with a
		phase singularity},}\ }\href {\doibase 10.1103/PhysRevLett.75.826} {\bibfield
{journal} {\bibinfo  {journal} {Phys. Rev. Lett.}\ }\textbf {\bibinfo
	{volume} {75}},\ \bibinfo {pages} {826--829} (\bibinfo {year}
{1995})}\BibitemShut {NoStop}%
\bibitem [{\citenamefont {Swartzlander}\ and\ \citenamefont
	{Hernandez-Aranda}(2007)}]{Swartzlander2007}%
\BibitemOpen
\bibfield  {author} {\bibinfo {author} {\bibfnamefont {Grover}\ \bibnamefont
		{Swartzlander}}\ and\ \bibinfo {author} {\bibfnamefont {Raul}\ \bibnamefont
		{Hernandez-Aranda}},\ }\bibfield  {title} {\enquote {\bibinfo {title}
		{{Optical Rankine Vortex and Anomalous Circulation of Light}},}\ }\href
{\doibase 10.1103/PhysRevLett.99.163901} {\bibfield  {journal} {\bibinfo
		{journal} {Phys. Rev. Lett.}\ }\textbf {\bibinfo {volume} {99}},\ \bibinfo
	{pages} {163901} (\bibinfo {year} {2007})}\BibitemShut {NoStop}%
\bibitem [{\citenamefont {Serabyn}\ \emph {et~al.}(2010)\citenamefont
	{Serabyn}, \citenamefont {Mawet},\ and\ \citenamefont
	{Burruss}}]{Serabyn2010}%
\BibitemOpen
\bibfield  {author} {\bibinfo {author} {\bibfnamefont {E}~\bibnamefont
		{Serabyn}}, \bibinfo {author} {\bibfnamefont {D}~\bibnamefont {Mawet}}, \
	and\ \bibinfo {author} {\bibfnamefont {R}~\bibnamefont {Burruss}},\
}\bibfield  {title} {\enquote {\bibinfo {title} {{An image of an exoplanet
			separated by two diffraction beamwidths from a star.}}}\ }\href {\doibase
10.1038/nature09007} {\bibfield  {journal} {\bibinfo  {journal} {Nature}\
}\textbf {\bibinfo {volume} {464}},\ \bibinfo {pages} {1018--20} (\bibinfo
{year} {2010})}\BibitemShut {NoStop}%
\bibitem [{\citenamefont {Berkhout}\ and\ \citenamefont
	{Beijersbergen}(2008)}]{Berkhout2008}%
\BibitemOpen
\bibfield  {author} {\bibinfo {author} {\bibfnamefont {Gregorius}\
		\bibnamefont {Berkhout}}\ and\ \bibinfo {author} {\bibfnamefont {Marco}\
		\bibnamefont {Beijersbergen}},\ }\bibfield  {title} {\enquote {\bibinfo
		{title} {{Method for Probing the Orbital Angular Momentum of Optical Vortices
				in Electromagnetic Waves from Astronomical Objects}},}\ }\href {\doibase
	10.1103/PhysRevLett.101.100801} {\bibfield  {journal} {\bibinfo  {journal}
		{Phys. Rev. Lett.}\ }\textbf {\bibinfo {volume} {101}},\ \bibinfo {pages}
	{100801} (\bibinfo {year} {2008})}\BibitemShut {NoStop}%
\bibitem [{\citenamefont {Wang}\ \emph {et~al.}(2012)\citenamefont {Wang},
	\citenamefont {Yang}, \citenamefont {Fazal}, \citenamefont {Ahmed},
	\citenamefont {Yan}, \citenamefont {Huang}, \citenamefont {Ren},
	\citenamefont {Yue}, \citenamefont {Dolinar}, \citenamefont {Tur},\ and\
	\citenamefont {Willner}}]{wangterabit}%
\BibitemOpen
\bibfield  {author} {\bibinfo {author} {\bibfnamefont {Jian}\ \bibnamefont
		{Wang}}, \bibinfo {author} {\bibfnamefont {Jeng-Yuan}\ \bibnamefont {Yang}},
	\bibinfo {author} {\bibfnamefont {Irfan~M}\ \bibnamefont {Fazal}}, \bibinfo
	{author} {\bibfnamefont {Nisar}\ \bibnamefont {Ahmed}}, \bibinfo {author}
	{\bibfnamefont {Yan}\ \bibnamefont {Yan}}, \bibinfo {author} {\bibfnamefont
		{Hao}\ \bibnamefont {Huang}}, \bibinfo {author} {\bibfnamefont {Yongxiong}\
		\bibnamefont {Ren}}, \bibinfo {author} {\bibfnamefont {Yang}\ \bibnamefont
		{Yue}}, \bibinfo {author} {\bibfnamefont {Samuel}\ \bibnamefont {Dolinar}},
	\bibinfo {author} {\bibfnamefont {Moshe}\ \bibnamefont {Tur}}, \ and\
	\bibinfo {author} {\bibfnamefont {Alan~E.}\ \bibnamefont {Willner}},\
}\bibfield  {title} {\enquote {\bibinfo {title} {Terabit free-space data
		transmission employing orbital angular momentum multiplexing},}\ }\href
{http://www.nature.com/nphoton/journal/v6/n7/full/nphoton.2012.138.html}
{\bibfield  {journal} {\bibinfo  {journal} {Nature Phot.}\ }\textbf {\bibinfo
		{volume} {6}},\ \bibinfo {pages} {488--496} (\bibinfo {year}
	{2012})}\BibitemShut {NoStop}%
\bibitem [{\citenamefont {Roychowdhury}\ \emph {et~al.}(2004)\citenamefont
	{Roychowdhury}, \citenamefont {Jaiswal},\ and\ \citenamefont
	{Singh}}]{Roychowdhury2004}%
\BibitemOpen
\bibfield  {author} {\bibinfo {author} {\bibfnamefont {Sanjoy}\ \bibnamefont
		{Roychowdhury}}, \bibinfo {author} {\bibfnamefont {Virendra~K.}\ \bibnamefont
		{Jaiswal}}, \ and\ \bibinfo {author} {\bibfnamefont {R.P.}\ \bibnamefont
		{Singh}},\ }\bibfield  {title} {\enquote {\bibinfo {title} {Implementing
			controlled not gate with optical vortex},}\ }\href {\doibase
	http://dx.doi.org/10.1016/j.optcom.2004.03.036} {\bibfield  {journal}
	{\bibinfo  {journal} {Optics Communications}\ }\textbf {\bibinfo {volume}
		{236}},\ \bibinfo {pages} {419 -- 424} (\bibinfo {year} {2004})}\BibitemShut
{NoStop}%
\bibitem [{\citenamefont {Verbeeck}\ \emph {et~al.}(2013)\citenamefont
	{Verbeeck}, \citenamefont {Tian},\ and\ \citenamefont {{Van
			Tendeloo}}}]{Verbeeck2013}%
\BibitemOpen
\bibfield  {author} {\bibinfo {author} {\bibfnamefont {Jo}~\bibnamefont
		{Verbeeck}}, \bibinfo {author} {\bibfnamefont {He}~\bibnamefont {Tian}}, \
	and\ \bibinfo {author} {\bibfnamefont {Gustaaf}\ \bibnamefont {{Van
				Tendeloo}}},\ }\bibfield  {title} {\enquote {\bibinfo {title} {{How to
				manipulate nanoparticles with an electron beam?}}}\ }\href {\doibase
	10.1002/adma.201204206} {\bibfield  {journal} {\bibinfo  {journal} {Adv.
			Mater.}\ }\textbf {\bibinfo {volume} {25}},\ \bibinfo {pages} {1114--7}
	(\bibinfo {year} {2013})}\BibitemShut {NoStop}%
\bibitem [{\citenamefont {Juchtmans}\ \emph {et~al.}(2015)\citenamefont
	{Juchtmans}, \citenamefont {B\'ech\'e}, \citenamefont {Abakumov},
	\citenamefont {Batuk},\ and\ \citenamefont {Verbeeck}}]{Juchtmans2015}%
\BibitemOpen
\bibfield  {author} {\bibinfo {author} {\bibfnamefont {Roeland}\ \bibnamefont
		{Juchtmans}}, \bibinfo {author} {\bibfnamefont {Armand}\ \bibnamefont
		{B\'ech\'e}}, \bibinfo {author} {\bibfnamefont {Artem}\ \bibnamefont
		{Abakumov}}, \bibinfo {author} {\bibfnamefont {Maria}\ \bibnamefont {Batuk}},
	\ and\ \bibinfo {author} {\bibfnamefont {Jo}~\bibnamefont {Verbeeck}},\
}\bibfield  {title} {\enquote {\bibinfo {title} {Using electron vortex beams
		to determine chirality of crystals in transmission electron microscopy},}\
}\href {\doibase 10.1103/PhysRevB.91.094112} {\bibfield  {journal} {\bibinfo
	{journal} {Phys. Rev. B}\ }\textbf {\bibinfo {volume} {91}},\ \bibinfo
{pages} {094112} (\bibinfo {year} {2015})}\BibitemShut {NoStop}%
\bibitem [{\citenamefont {Schattschneider}\ \emph {et~al.}(2014)\citenamefont
	{Schattschneider}, \citenamefont {L{\"{o}}ffler}, \citenamefont
	{St{\"{o}}ger-Pollach},\ and\ \citenamefont
	{Verbeeck}}]{Schattschneider2014}%
\BibitemOpen
\bibfield  {author} {\bibinfo {author} {\bibfnamefont {P}~\bibnamefont
		{Schattschneider}}, \bibinfo {author} {\bibfnamefont {S}~\bibnamefont
		{L{\"{o}}ffler}}, \bibinfo {author} {\bibfnamefont {M}~\bibnamefont
		{St{\"{o}}ger-Pollach}}, \ and\ \bibinfo {author} {\bibfnamefont
		{J}~\bibnamefont {Verbeeck}},\ }\bibfield  {title} {\enquote {\bibinfo
		{title} {{Is magnetic chiral dichroism feasible with electron vortices?}}}\
}\href {\doibase 10.1016/j.ultramic.2013.07.012} {\bibfield  {journal}
{\bibinfo  {journal} {Ultramicroscopy}\ }\textbf {\bibinfo {volume} {136}},\
\bibinfo {pages} {81--5} (\bibinfo {year} {2014})}\BibitemShut {NoStop}%
\bibitem [{\citenamefont {Berkhout}\ and\ \citenamefont
	{Beijersbergen}(2009)}]{Berkhout2009}%
\BibitemOpen
\bibfield  {author} {\bibinfo {author} {\bibfnamefont {G~C~G}\ \bibnamefont
		{Berkhout}}\ and\ \bibinfo {author} {\bibfnamefont {M~W}\ \bibnamefont
		{Beijersbergen}},\ }\bibfield  {title} {\enquote {\bibinfo {title} {{Using a
				multipoint interferometer to measure the orbital angular momentum of light in
				astrophysics}},}\ }\href {\doibase 10.1088/1464-4258/11/9/094021} {\bibfield
	{journal} {\bibinfo  {journal} {Journal of Optics A: Pure and Applied
			Optics}\ }\textbf {\bibinfo {volume} {11}},\ \bibinfo {pages} {094021}
	(\bibinfo {year} {2009})}\BibitemShut {NoStop}%
\bibitem [{\citenamefont {Liu}(2015)}]{Liu2015}%
\BibitemOpen
\bibfield  {author} {\bibinfo {author} {\bibfnamefont {Man}\ \bibnamefont
		{Liu}},\ }\bibfield  {title} {\enquote {\bibinfo {title} {{Measuring the
				fractional and integral topological charges of the optical vortex beams using
				a diffraction grating}},}\ }\href {\doibase 10.1016/j.ijleo.2015.09.004}
{\bibfield  {journal} {\bibinfo  {journal} {Optik - International Journal for
			Light and Electron Optics}\ } (\bibinfo {year} {2015}),\
	10.1016/j.ijleo.2015.09.004}\BibitemShut {NoStop}%
\bibitem [{\citenamefont {Yongxin}\ \emph {et~al.}(2011)\citenamefont
	{Yongxin}, \citenamefont {Hua}, \citenamefont {Jixiong},\ and\ \citenamefont
	{Baida}}]{Yongxin2011}%
\BibitemOpen
\bibfield  {author} {\bibinfo {author} {\bibfnamefont {Liu}\ \bibnamefont
		{Yongxin}}, \bibinfo {author} {\bibfnamefont {Tao}\ \bibnamefont {Hua}},
	\bibinfo {author} {\bibfnamefont {Pu}~\bibnamefont {Jixiong}}, \ and\
	\bibinfo {author} {\bibfnamefont {L\"{u}}\ \bibnamefont {Baida}},\ }\bibfield
{title} {\enquote {\bibinfo {title} {{Detecting the topological charge of
				vortex beams using an annular triangle aperture}},}\ }\href {\doibase
	10.1016/j.optlastec.2011.03.015} {\bibfield  {journal} {\bibinfo  {journal}
		{Optics \& Laser Technology}\ }\textbf {\bibinfo {volume} {43}},\ \bibinfo
	{pages} {1233--1236} (\bibinfo {year} {2011})}\BibitemShut {NoStop}%
\bibitem [{\citenamefont {Guzzinati}\ \emph {et~al.}(2014)\citenamefont
	{Guzzinati}, \citenamefont {Clark}, \citenamefont {B\'{e}ch\'{e}},\ and\
	\citenamefont {Verbeeck}}]{Guzzinati2014}%
\BibitemOpen
\bibfield  {author} {\bibinfo {author} {\bibfnamefont {Giulio}\ \bibnamefont
		{Guzzinati}}, \bibinfo {author} {\bibfnamefont {Laura}\ \bibnamefont
		{Clark}}, \bibinfo {author} {\bibfnamefont {Armand}\ \bibnamefont
		{B\'{e}ch\'{e}}}, \ and\ \bibinfo {author} {\bibfnamefont {Jo}~\bibnamefont
		{Verbeeck}},\ }\bibfield  {title} {\enquote {\bibinfo {title} {{Measuring the
				orbital angular momentum of electron beams}},}\ }\href {\doibase
	10.1103/PhysRevA.89.025803} {\bibfield  {journal} {\bibinfo  {journal}
		{Physical Review A}\ }\textbf {\bibinfo {volume} {89}},\ \bibinfo {pages}
	{025803} (\bibinfo {year} {2014})},\ \Eprint {http://arxiv.org/abs/1401.7211}
{1401.7211} \BibitemShut {NoStop}%
\bibitem [{\citenamefont {Dudley}\ \emph {et~al.}(2012)\citenamefont {Dudley},
	\citenamefont {Litvin},\ and\ \citenamefont {Forbes}}]{Dudley2012}%
\BibitemOpen
\bibfield  {author} {\bibinfo {author} {\bibfnamefont {Angela}\ \bibnamefont
		{Dudley}}, \bibinfo {author} {\bibfnamefont {Igor~a.}\ \bibnamefont
		{Litvin}}, \ and\ \bibinfo {author} {\bibfnamefont {Andrew}\ \bibnamefont
		{Forbes}},\ }\bibfield  {title} {\enquote {\bibinfo {title} {{Quantitative
				measurement of the orbital angular momentum density of light}},}\ }\href
{\doibase 10.1364/AO.51.000823} {\bibfield  {journal} {\bibinfo  {journal}
		{Applied Optics}\ }\textbf {\bibinfo {volume} {51}},\ \bibinfo {pages} {823}
	(\bibinfo {year} {2012})}\BibitemShut {NoStop}%
\bibitem [{\citenamefont {Schulze}\ \emph {et~al.}(2013)\citenamefont
	{Schulze}, \citenamefont {Dudley}, \citenamefont {Flamm}, \citenamefont
	{Duparr\'{e}},\ and\ \citenamefont {Forbes}}]{Schulze2013}%
\BibitemOpen
\bibfield  {author} {\bibinfo {author} {\bibfnamefont {Christian}\
		\bibnamefont {Schulze}}, \bibinfo {author} {\bibfnamefont {Angela}\
		\bibnamefont {Dudley}}, \bibinfo {author} {\bibfnamefont {Daniel}\
		\bibnamefont {Flamm}}, \bibinfo {author} {\bibfnamefont {Michael}\
		\bibnamefont {Duparr\'{e}}}, \ and\ \bibinfo {author} {\bibfnamefont
		{Andrew}\ \bibnamefont {Forbes}},\ }\bibfield  {title} {\enquote {\bibinfo
		{title} {{Measurement of the orbital angular momentum density of light by
				modal decomposition}},}\ }\href {\doibase 10.1088/1367-2630/15/7/073025}
{\bibfield  {journal} {\bibinfo  {journal} {New Journal of Physics}\ }\textbf
	{\bibinfo {volume} {15}} (\bibinfo {year} {2013}),\
	10.1088/1367-2630/15/7/073025}\BibitemShut {NoStop}%
\bibitem [{\citenamefont {Q.Wang}\ \emph {et~al.}(2008)\citenamefont {Q.Wang},
	\citenamefont {O.Ronneberger},\ and\ \citenamefont {H.Burkhardt}}]{Wang}%
\BibitemOpen
\bibfield  {author} {\bibinfo {author} {\bibnamefont {Q.Wang}}, \bibinfo
	{author} {\bibnamefont {O.Ronneberger}}, \ and\ \bibinfo {author}
	{\bibnamefont {H.Burkhardt}},\ }\href
{http://lmb.informatik.uni-freiburg.de/papers/download/wa_report01_08.pdf}
{\emph {\bibinfo {title} {Fourier Analysis in Polar and Spherical
			Coordinates}}},\ \bibinfo {type} {Tech. Rep.}\ \bibinfo {number} {1}\
(\bibinfo  {institution} {IIF-LMB, Computer Science Department, University of
	Freiburg},\ \bibinfo {year} {2008})\BibitemShut {NoStop}%
\bibitem [{\citenamefont {F\"urhapter}\ \emph {et~al.}(2005)\citenamefont
	{F\"urhapter}, \citenamefont {Jesacher}, \citenamefont {Bernet},\ and\
	\citenamefont {Ritsch-Marte}}]{Furhapter2005}%
\BibitemOpen
\bibfield  {author} {\bibinfo {author} {\bibfnamefont {Severin}\ \bibnamefont
		{F\"urhapter}}, \bibinfo {author} {\bibfnamefont {Alexander}\ \bibnamefont
		{Jesacher}}, \bibinfo {author} {\bibfnamefont {Stefan}\ \bibnamefont
		{Bernet}}, \ and\ \bibinfo {author} {\bibfnamefont {Monika}\ \bibnamefont
		{Ritsch-Marte}},\ }\bibfield  {title} {\enquote {\bibinfo {title} {{Spiral
				phase contrast imaging in microscopy}},}\ }\href {\doibase
	10.1364/OPEX.13.000689} {\bibfield  {journal} {\bibinfo  {journal} {Optics
			Express}\ }\textbf {\bibinfo {volume} {13}},\ \bibinfo {pages} {689}
	(\bibinfo {year} {2005})}\BibitemShut {NoStop}%
\bibitem [{\citenamefont {Tonomura}(1987)}]{Tonomura1987}%
\BibitemOpen
\bibfield  {author} {\bibinfo {author} {\bibfnamefont {A}~\bibnamefont
		{Tonomura}},\ }\bibfield  {title} {\enquote {\bibinfo {title} {{Applications
				of electron holography}},}\ }\href
{http://journals.aps.org/rmp/abstract/10.1103/RevModPhys.59.639} {\bibfield
	{journal} {\bibinfo  {journal} {Reviews of modern physics}\ } (\bibinfo
	{year} {1987})}\BibitemShut {NoStop}%
\bibitem [{\citenamefont {Verbeeck}\ \emph {et~al.}(2015)\citenamefont
	{Verbeeck}, \citenamefont {B\'ech\'e}, \citenamefont {Guzzinati},
	\citenamefont {Clarck}, \citenamefont {Juchtmans}, \citenamefont
	{Van~Boxem},\ and\ \citenamefont {Van~Tendeloo}}]{Verbeeck2015}%
\BibitemOpen
\bibfield  {author} {\bibinfo {author} {\bibfnamefont {Jo}~\bibnamefont
		{Verbeeck}}, \bibinfo {author} {\bibfnamefont {Armand}\ \bibnamefont
		{B\'ech\'e}}, \bibinfo {author} {\bibfnamefont {Giulio}\ \bibnamefont
		{Guzzinati}}, \bibinfo {author} {\bibfnamefont {Laura}\ \bibnamefont
		{Clarck}}, \bibinfo {author} {\bibfnamefont {Roeland}\ \bibnamefont
		{Juchtmans}}, \bibinfo {author} {\bibfnamefont {Ruben}\ \bibnamefont
		{Van~Boxem}}, \ and\ \bibinfo {author} {\bibfnamefont {Gustaaf}\ \bibnamefont
		{Van~Tendeloo}},\ }\bibfield  {title} {\enquote {\bibinfo {title}
		{Construction of a programmable electron vortex phase plate},}\ }in\ \href
{http://epub.uni-regensburg.de/32876/1/MC2015_Proceedings.pdf} {\emph
	{\bibinfo {booktitle} {MC 2015 Proceedings}}}\ (\bibinfo {organization} {DGE
	- German Society for Electon Microscopy},\ \bibinfo {year} {2015})\ pp.\
\bibinfo {pages} {582--583}\BibitemShut {NoStop}%
\end{thebibliography}
\end{document}